\documentclass[twocolumn,aps,superscriptaddress,10pt,dvips,showpacs,showkeys,prl]{revtex4}
\usepackage[utf8]{inputenc}
\usepackage{graphicx}
\usepackage{setspace}
\usepackage{units}
\usepackage{xspace}
\usepackage{color}

\begin{document}

\title{Effects of non-uniform Mn distribution in (Ga,Mn)As} 
\author{I. Ulfat} 
\email{intikhab.ulfat@chalmers.se}
\affiliation{Department of Experimental Physics, Chalmers University
  of Technology, SE-412 96 G\"oteborg, Sweden}
\author{J. Kanski} 
\email{janusz.kanski@chalmers.se}
\affiliation{Department of Experimental Physics, Chalmers University
  of Technology, SE-412 96 G\"oteborg, Sweden}
\author{L. Ilver} 
\email{lars.ilver@chalmers.se}
\affiliation{Department of Experimental Physics, Chalmers University
  of Technology, SE-412 96 G\"oteborg, Sweden}
\author{J. Sadowski} 
\email{janusz.sadowski@maxlab.lu.se>}
\affiliation{MAX-lab, Lund University,
 SE-221 00 Lund, Sweden}
\author{K. Karlsson}
\email{krister.karlsson@his.se>}
\affiliation{Department of Life Sciences, University of Sk\"ovde,
 SE-541 28 Sk\"ovde, Sweden}
\author{A.~Ernst}
\email{aernst@mpi-halle.de} 
\affiliation{Max-Planck-Institut f\"ur
  Mikrostrukturphysik, Weinberg 2, D-06120 Halle, Germany}
\date{\today}
\begin{abstract}
  Resonant in situ photoemission from Mn $3d$ states in
  Ga$_{1-x}$Mn$_{x}$As is reported for Mn concentrations down to very
  dilute limit of \unit[0.1]{at\%}. The properties of the peak at the
  valence-band maximum reveal an effective interaction between Mn 3d
  states for concentration as low as 2.5\%.  Concentration-dependent
  spectral features are analyzed on the basis of first-principles
  calculations for systems with selected impurity positions as well as
  for random alloys.
\end{abstract}
\maketitle
 
In the quest for magnetic semiconductors with potential use in
spintronics, (Ga,Mn)As has emerged as the prototype system with
documented spin-polarization of the electron states. Even though the
Curie temperature of (Ga,Mn)As is still too low for practical
implementations, the transport properties can be controlled to the
extent that device structures based on the unusual magnetic
characteristics can be explored. In this perspective
it is remarkable that the electronic structure of (Ga,Mn)As remains
poorly characterized. It is generally agreed that the ferromagnetic
state is due to Mn atoms in substitutional Ga sites. From electron
spin resonance (ESR) and optical data it is known that individual Mn
atoms occur in either an ionized $3d^5$ or a neutral ($3d^5$+hole)
state, the latter being the reason for the $p$-type behaviour with the
acceptor level at \unit[113]{meV} above the valence band maximum
(VBM)~\cite{Schneider1987,Twardowski1999}. In systems with higher
Mn concentrations the ESR data are found to be strongly
broadened~\cite{Twardowski1999} and the situation is less clear.
Interestingly, the broadening is observed in epitaxial layers with
relatively low Mn concentration (\unit[0.5]{at\%}), where one would
expect to have a system of well separated impurities. The broadening
has been tentatively ascribed to effects of demagnetizing
field~\cite{Twardowski1999}.

Photoelectron spectroscopy is the most direct method for probing the electron structure. An issue of concern in this context is the
intrinsic surface sensitivity of this technique. This, in combination
with the metastable character of (Ga,Mn)As with concentrations of
interest (in the range of \unit[1]{\%}) prohibits standard surface
preparation involving ion beam surface cleaning and annealing at
sufficiently high temperatures (above \unit[300]{$^{\circ}$C}) to
restore the surface order. Although this is obvious, it is remarkable that only a few {\em in
  situ} studies are available so far.  Even more surprising,
some of the ”generally accepted” details about the electronic
structure are derived from studies involving such treatment even though
they are in conflict with results from {\em in situ} studies.
Specifically it has been demonstrated that the energy of the main Mn
$3d$–induced valence band structure is shifted by about 1 eV towards
higher binding energy by such treatment~\cite{Adell2004}, yet the most
quoted value is that derived from {\em ex situ} treated
samples~\cite{Rader2009}. 

A particularly important issue in the context of magnetism in
(Ga,Mn)As is the nature of electron states mediating the ferromagnetic
coupling. The so far most successful description of ferromagnetism in
(Ga,Mn)As has been based on spin polarization of holes in the GaAs
valence band~\cite{Dietl2000}.  More recently this picture has been
questioned, since experiments showed that the ferromagnetic state was
retained even in the absence of valence band holes ~\cite{Scarpulla2005}.  

The present letter reports a detailed in situ resonant photoemission
study for samples with different Mn concentration in the range
\unit[0.1]{\%} to \unit[6]{\%}.  It is emphasized that these are the
first photoemission data for such a very dilute limit.  We focus our
attention on the 3d Mn peak at the valence band maximum since the
states in this energy region are crucial for the establishing of long
range magnetic order. We show that at Mn concentrations as low as
2.5\% there is clear sign of effective interaction between Mn 3d
states.  The analysis of the spectra is based on two types of
first-principles calculations: for samples with selected impurity
positions and for random alloys.

The photoemission data were obtained at the undulator beamline I511 at
the Swedish synchrotron radiation facility MAX-lab.  The samples were
prepared in a local molecular beam epitaxy~(MBE) system, and were
transferred to the photoemission station in a portable UHV chamber
without being exposed to atmosphere. The Mn concentrations were
determined by means of reflection high-energy electron
diffraction~(RHEED) oscillations, as described
earlier~\cite{Sadowski2000}. The RHEED oscillations were also used for
defining a secondary Mn concentration scale based on the Mn $2p_{3/2}$
absorption spectra, which was used in cases where clear oscillations
were not observed. It is important to stress that the (Ga,Mn)As
samples were not subject to any post-growth treatment. The surface
cleanliness was checked by survey spectra recorded at \unit[1000]{eV}
photon energy. All samples were perfectly free from surface
contamination.

\begin{figure}[!bth]
\begin{center}
\includegraphics[width=0.95\columnwidth]{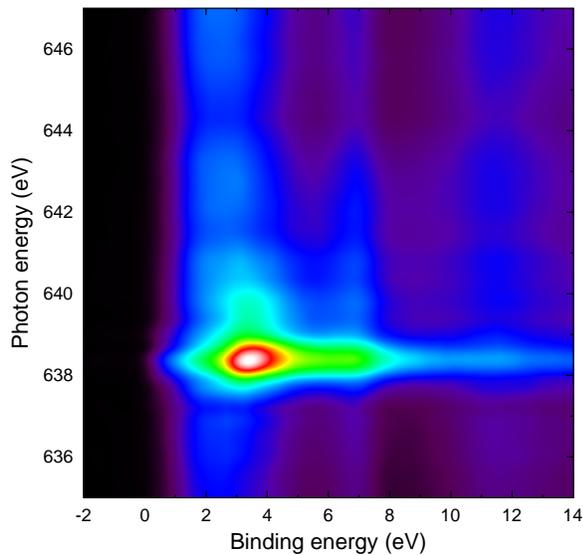}
\end{center}
%\vskip -1 cm
\caption{Valence band photoemission intensity contour plot above the Mn $2p_{3/2}$ 
excitation. Note the absence of any structures along the diagonal that would indicate 
incoherent Auger emission.}
\label{fig:1}
\end{figure}

Valence band photoemission from dilute systems like (Ga,Mn)As is
normally dominated by the host material states (GaAs in the present
case).  However, emission from impurity states can be enhanced
selectively under resonant conditions, i.e. when the photoemission
process involves an intermediate core hole excitation. In the present
system a strong enhancement is expected just above the Mn $2p$
excitation threshold, due to a large $2p-3d$ excitation matrix element
in combination with a localized $2p^5 3d^{n+1}$ intermediate state.
This is indeed verified in the intensity contour plot in
Fig.~\ref{fig:1}. The complete lack of structures along the diagonal
in this graph (i.e. structures reflecting Auger decay) shows that the
enhancement is predominantly caused by resonant photoemission. This is
in contrast to the results in ref.~\onlinecite{Rader2004}, where about
50\% of the spectral intensity was concluded to be due to incoherent
Auger transitions.  Since the incoherent decay becomes prominent with
Mn clustering~\cite{Adell2004,Adell2011}, this observation together
with the deeper position of the main $3d$ peak, indicates that the
sample in ref.~\onlinecite{Rader2004} contained Mn clusters, likely
due to post-growth treatment.

\begin{figure}[!bth]
\begin{center}
\includegraphics[width=0.95\columnwidth]{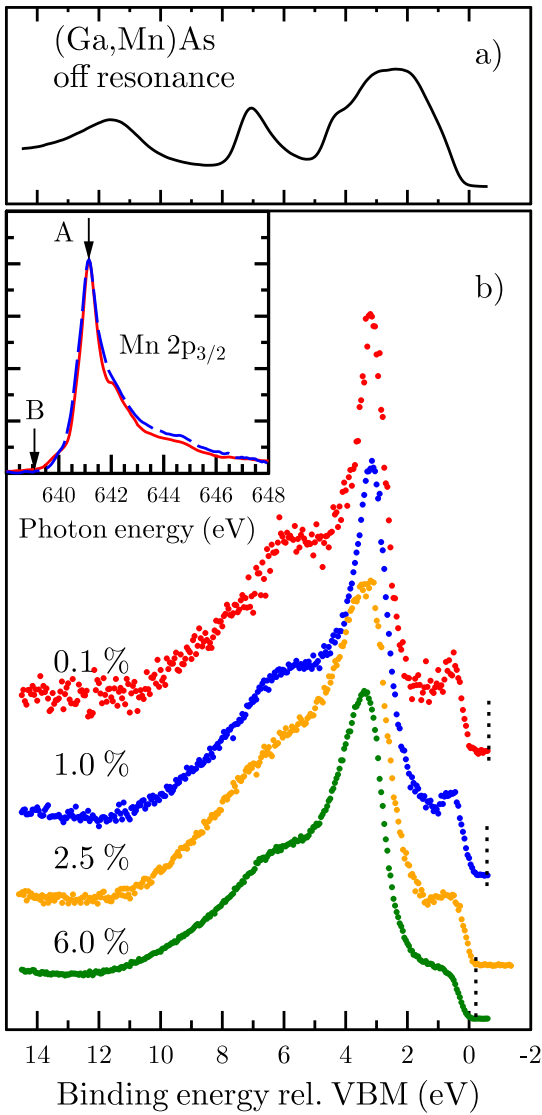}
\end{center}
\caption{a) Valence band spectrum of (Ga,Mn)As with \unit[6]{\%}) Mn,
  excited with \unit[639]{eV} photons (i.e. below resonance). b)
  Difference curves between spectra obtained on- and off resonance for
  three Mn concentrations. The intensities are normalized at the high
  binding energy region, and aligned in energy with respect to the
  valence band maxima. For each curve the Fermi level position is
  indicated with a vertical dashed line. The inset shows the XAS
  spectra from samples with \unit[0.3]{\%}) and \unit[6]{\%}) Mn
  (solid and dashed curves, respectively.) }
\label{fig:2}
\end{figure}

The resonant valence band emission (reflecting the Mn $3d$-states) was
obtained from the difference between spectra recorded with photon
energies on- and off-resonance, as indicated in the inset in
Fig.~\ref{fig:2}). The main advantage of this procedure, as compared with that in ref. 8, is
that it avoids artefacts not related to the resonant process, e.g.
different surface states on different samples.

In situations with very low impurity concentrations, the spectra recorded    
on- and off-resonance are of course very similar, and the extraction of 
difference spectra becomes a quite delicate procedure. In such situations 
it is extremely important that the spectra are properly normalized and 
well aligned with respect to binding energy. We achieved this by extending 
the recorded spectra to cover the Ga $3d$ peak, and used this peak for normalizing 
as well as aligning spectra obtained at different photon energies. The alignment 
was done with an accuracy of \unit[1]{meV}, and the procedure was checked 
by the absence of any systematic structures in the Ga $3d$ difference spectra.
For the most dilute cases it turned out that thermal effects due to varying 
monochromator heating had to be considered. To minimize such effects, the 
recording time for each spectrum was reduced to about 1 minute. To achieve 
reasonable statistics, a number (typically 50-100) of such difference spectra were 
added after proper alignment.

In Fig.~\ref{fig:2} we show a set of valence band difference curves
obtained for samples with different Mn concentrations. Our
high-concentration data resemble to some extent those presented in
ref. 8, but as will be pointed out below, there are also significant
differences.  The analysis shows that the main Mn
$3d$–induced valence band structure is observed around
\unit[3.0-3.3]{eV} below the VBM. With increasing Mn concentration
this peak broadens and shifts by 0.2-0.3 eV towards higher binding
energy.
Another Mn induced structure is observed close to VBM. At low
concentrations it is peak-shaped, but broadens and becomes
shoulder-like for high concentrations. The electron states in this
energy region are crucial for establishing long-range magnetic order
and demand a careful study. Our experiments are performed for a broad range
of Mn concentrations, including the very dilute limit of \unit[0.1]{\%}, and provides new
possibilities for investigations  of the interaction between distant impurity
atoms.

To elucidate the experimental findings we performed an extensive
first-principles study using a self-consistent Green function
Korringa-Kohn-Rostoker~(KKR) method implemented within the multiple
scattering theory~\cite{Luders2001}. The calculations were performed
within the density functional theory in the local spin density
approximation~(LSDA).  Two types of calculations have been performed.
First, we considered individual Mn impurities in GaAs by placing them
at different positions inside a large fragment of the semiconductor.
Second, the coherent potential approximation~(CPA) was used to
consider random alloys with different Mn concentrations.  According to
our CPA calculations (not displayed) the main Mn $3d$ peak observed in the experiment
around \unit[3.0-3.3]{eV} can be clearly identified with Mn atoms in
substitutional Ga sites. In the calculations this peak is found at
\unit[2.9]{eV} below the Fermi level. This result is in a very good
agreement with other first-principles calculations performed within
the local density approximation. As already mentioned, the position of
the main Mn $3d$ peak is an issue of controversy, and it is worth
stressing that the consistency between the present experimental and
theoretical results is in conflict with earlier data locating the peak
around \unit[4.5]{eV} below VBM ~\cite{Rader2004,Okabayashi1999}.

\begin{figure}[t]
\begin{center}
\includegraphics[width=0.95\columnwidth]{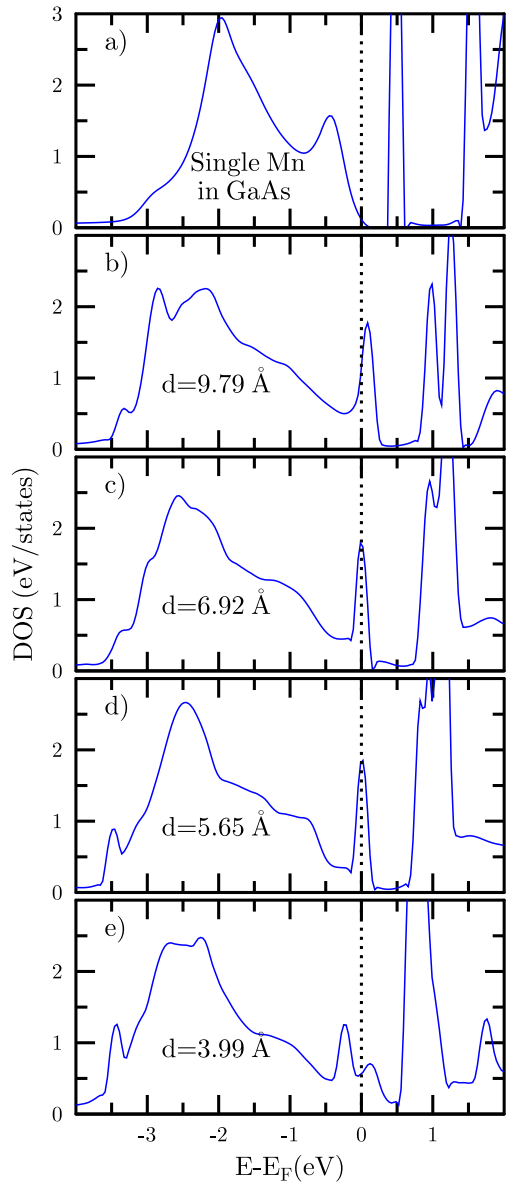}
\end{center}
\caption{The density of states of Mn impurities in GaAs calculated
  for various Mn-Mn distances: (a) single impurity; (b-e) two
  impurities separated by \unit[9.79]{\AA}, \unit[6.92]{\AA},
  \unit[5.65]{\AA}, and \unit[3.99]{\AA} respectively.}
\label{fig:3}
\end{figure}

\begin{figure}[t]
\begin{center}
\includegraphics[width=0.95\columnwidth]{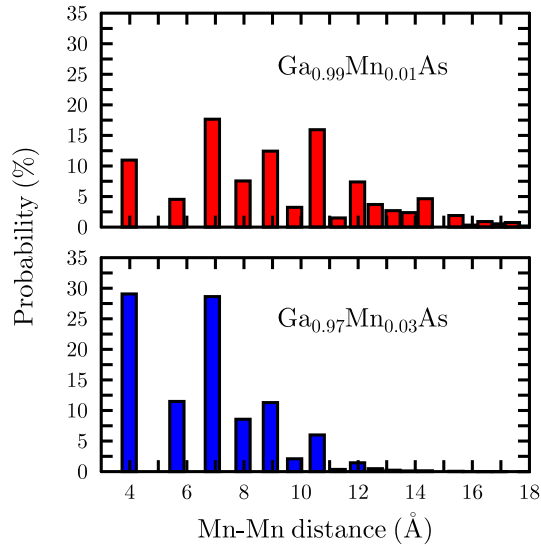}
\end{center}
\caption{The distribution of distances to nearest neighbors Mn
atoms for concentrations corresponding to  Ga$_{0.99}$Mn$_{0.01}$As~(a) and  Ga$_{0.97}$Mn$_{0.03}$As~(b).}
\label{fig:4}
\end{figure}

In order to address the concentration dependence we begin with calculations of
the density of states of an isolated Mn impurity at a Ga site. We
obtain a narrow peak above the Fermi level~(Fig. \ref{fig:3}a). These
empty states reflect the acceptor nature of the Mn$_{Ga}$ impurity.
The integral of the DOS above the Fermi level shows that, as expected,
there is exactly one hole per Mn atom. 

The CPA calculation for the 0.1\% concentration of Mn$_{Ga}$
impurities gives the result very similar to those for an isolated
impurity.  Since in the low concentration limit
the CPA calculation should reproduce the results for the isolated
impurity, the agreement of two different approaches demonstrates the
stability of the numerical procedures.

If we place another Mn$_{Ga}$ impurity at a distance of \unit[5.65]{\AA}
or more from the first one, the narrow peak is shifted close to the
VBM and becomes partly occupied, although the shape of the Mn DOS
remains almost unchanged. The two Mn atoms do not noticeably influence
each other. Merely in the case of the nearest next neighbors at the
distance of \unit[3.99]{\AA} the peak becomes dispersive, indicating a
strong hybridization between Mn $3d$-states.

According to our experimental findings~(Fig.\ref{fig:2}) the peak close to
the VBM is broadened for a concentration of 2.5\%.  In the case of
uniformly distributed Mn impurities in Ga$_{0.975}$Mn$_{0.025}$As, the
distance between Mn atomic positions is around \unit[12]{\AA}. At such
distances the Mn-Mn hybridization is insignificant and will not broaden the
peak close the VBM (see Fig.~\ref{fig:3}). Therefore, we have reasons to believe
that the distribution of Mn impurities is not uniform and strongly depends on
the Mn concentration. To justify this, we have performed
Monte Carlo simulations with 10.000 Mn atoms randomly distributed in a
fcc matrix at different concentrations. The results of our simulations
for 1\% and 3\% are shown in Fig.~\ref{fig:4}. The most important and striking
result is the high proportion of nearest neighbors at high Mn
concentrations. In the case of the 3\% alloy, 30\% of Mn atoms have
another Mn in an adjacent site at \unit[3.99]{\AA}. At very low
concentration (0.1\%) the distribution is peaked halfway to the
nominal mean distance of \unit[36]{\AA} (is not shown here). 

The finding that the variation of the spectral features has very
strong dependence on the impurity concentration is relevant for the
discussion of the ferromagnetism of GaMnAs. Indeed, in the mediating
of the interatomic exchange interaction the states close to the VBM play
the most important role. The substantial broadening of the spectral
features in this energy reagion already for relatively low Mn
concentration explains the high Curie temperature in such samples.

In conclusion, the observed broadening of the Mn $3d$-states in
Ga$_{1-x}$Mn$_{x}$As for $x>1\%$~ can be explained by a highly
non-uniform local distribution of Mn-impurities in GaAs and the strong
$d$-$d$ hybridization between nearest Mn atoms. Our combined
experimental-theoretical approach allows to detect and interpret the
features in the Mn $3d$ related DOS that appear most clear in the
singular impurity limit and broaden very quickly with increasing Mn
concentration. 

\bibliography{./GaAsMn}
\bibliographystyle{apsrev}
\end{document}